\documentclass[aps,prb,twocolumn,showpacs,floatfix,superscriptaddress,amssymb,
amsmath]{revtex4}
\usepackage{graphicx}
\usepackage{dcolumn}
\usepackage{bm}
\usepackage{amsmath}

\begin{document}
\title{Kondo regime in triangular arrangements of quantum dots:  Molecular orbitals, interference and contact effects}
\author{E. Vernek}
\affiliation{Department of Physics and Astronomy, and Nanoscale
and Quantum Phenomena Institute, \\Ohio University, Athens, Ohio
45701-2979}
\affiliation{Instituto de F\'isica - Universidade Federal de Uberl\^andia - Uberl\^andia, 
MG 38400-902 - Brazil}
\author{C. A. B\"usser}
\affiliation{Department of Physics and Astronomy, and Nanoscale
and Quantum Phenomena Institute, \\Ohio University, Athens, Ohio
45701-2979}
\affiliation{Department of Physics, Oakland University, Rochester,
MI 48309, USA}
\author{G. B. Martins}
\email[Corresponding author:]{martins@oakland.edu}
\affiliation{Department of Physics, Oakland University, Rochester,
MI 48309, USA}
\author{E. V. Anda}
\affiliation{Departamento de F\'{\i}sica, Pontif\'{\i}cia
Universidade Cat\'olica, Rio de Janeiro-RJ, Brazil}
\author{N. Sandler}
\affiliation{Department of Physics and Astronomy, and Nanoscale
and Quantum Phenomena Institute, \\Ohio University, Athens, Ohio
45701-2979}
\author{S. E. Ulloa}
\affiliation{Department of Physics and Astronomy, and Nanoscale
and Quantum Phenomena Institute, \\Ohio University, Athens, Ohio
45701-2979}
\date{\today}
\begin{abstract}
Transport properties of an interacting triple quantum dot system coupled to three leads
in a triangular geometry has been studied in the Kondo regime. Applying mean-field 
finite-U slave boson and embedded cluster approximations 
to the calculation of transport properties unveils a set of 
rich features associated to the high symmetry of this system.  Results using both calculation techniques yield excellent overall agreement and provide additional
insights into the physical behavior of this interesting geometry. In the case when 
just two current leads are connected to the three-dot system, interference effects between degenerate molecular orbitals are found to strongly affect the overall conductance.  An $S=1$ Kondo effect is also shown to appear for the perfect equilateral triangle symmetry.  
The introduction of a third current lead results in an `amplitude leakage' phenomenon, 
akin to that appearing in beam splitters, which alters the interference effects and the overall conductance through the system.
\end{abstract}

\maketitle
\section{Introduction}
One of the most important and exciting aspects of molecular physics nowadays is the
study of electronic transport properties of natural and/or fabricated structures at
the nanoscopic scale. In nature, molecules can couple to the external environment
(electron reservoirs) through extended orbitals, which permit conduction electrons to
hop in and out of the molecule. Fabricated molecules can be made by coupled
quantum dots (QDs) with discrete energy
levels. \cite{kouwenhoven,bednarek,heath} Depending upon the strength of the coupling between the QDs,
they can behave as a molecule with extended
orbitals, which can further couple to external electron reservoirs. The confinement of electrons inside this artificial `molecule' produces strong
Coulomb interactions,\cite{ramirez} which may give rise under suitable conditions to Kondo physics for temperatures
below a characteristic crossover temperature, the Kondo
temperature, $T_K$.\cite{hewson,pustilnik,izumida} In the
simplest picture of this regime, for $T \ll T_{\rm K}$, the system forms a singlet state,
created by the screening of the localized spin by the conduction electrons in the external reservoir. 
Since its first observation in QDs in 1998,\cite{david-nature} the attention generated by the Kondo
effect in these structures has led to an explosion in experiments and theory. 
For example, multiple QD systems have become platforms for the theoretical and experimental development of sophisticated 
arrangements in order to access the rich phenomenology of the Kondo problem, 
including non-Fermi-liquid behavior and quantum critical points.\cite{Potok}

In this paper, we study the transport properties of a triple quantum dot 
(TQD) system in the `molecular regime' [with strong interdot couplings; see Fig.\ \ref{fig1}(a)] in two distinct situations:
Firstly, just two QDs are attached to independent electron reservoirs.
Secondly, each QD is connected to an independent electron reservoir. In
the latter case, we focus our attention on the conductance of the system through two of the
three terminals (the same two used to measure conductance in the first case). We are
particularly interested in understanding interference effects,
especially the role played by the third lead in the propagation of electrons along the
different trajectories. 

Despite significant advances in the understanding of Kondo physics in double\cite{meir,vernek,cornaglia,jeong,busser} and
triple\cite{zitko2,jiang,lobos,kuzmenko1,zitko1,zitko1b} QD structures made in the last few years, there are still important aspects of the problem which deserve to be studied in detail.
For example, based on a suggestion by Zarand {\it et al.},\cite{zarand} one may 
ask if an SU(4) Kondo regime may be experimentally attained in a TQD geometry. In addition, 
the unprecedented control of parameters in these multi-dot structures opens the possibility of 
observing quantum critical points and their associated non-Fermi-liquid ground states.
Although many of these have 
been theoretically identified, the very demanding experimental constraints required have resulted in only a few successful experimental realizations.\cite{Potok}
Further motivation to study TQD systems comes from the proposal
by Saraga and Loss that these structures could be used to produce spatially separated
currents of spin-entangled electrons.\cite{loss} Experimentally, however, only few groups have reported work in these systems.\cite{gaudreau,vidan,rogge} Most
of these studies have been in the Coulomb blockade regime, and 
one of the works reports that a TQD device can act as a molecular rectifier.\cite{vidan} 

\v{Z}itko and Bon\u{c}a\cite{zitko1} have recently studied theoretically a 
TQD system connected in series to two leads. 
They have found that for a certain range of inter-dot hopping parameters the system crosses 
over from a Fermi-liquid to a non-Fermi-liquid regime in a wide interval of
temperatures. In a subsequent paper,\cite{zitko1b} using the numerical renormalization group 
(NRG), these authors analyze a large number of 
phases for a system similar to the one we discuss in this paper.  Notice, nevertheless, that there are important differences between their system and ours: The majority of the phases analyzed in detail 
in Ref.\ \onlinecite{zitko1b} use a Kondo Hamiltonian for the dots. In that case, a crossover was predicted
between the two-impurity and the two-channel Kondo-model non-Fermi-liquid fixed points. 
Their analysis of Anderson-impurity QDs brings their work closer to ours. However, they restrict 
their study mainly to regimes where the inter-dot hoppings are considerably smaller than the coupling to the 
leads, which is exactly the opposite regime we treat in our work. In addition, they only analyze results close to half-filling, 
while we consider all fillings. Finally, and more importantly, they did not analyze the very important influence of the third contact,  
which is one of the important results of the work we report here.

The physics of this arrangement of QDs has also been the subject of other theoretical
works.\cite{lobos,ingersent1,kuzmenko,kikoin,avishai,sakano,emary}  
In particular, a situation where the system may present interesting, but rather complicated behavior, is in the
fully symmetric case, {\it i.e.}, when all inter-dot hoppings are the same and 
each QD is equally connected to an independent conducting band (in that case, the system has equilateral triangle
symmetry). It is reasonable then, if the inter-dot hoppings are much smaller than the 
intra-dot Coulomb repulsion, to expect the TQD system to present a spin frustrated regime, as anti-ferromagnetic arrangement between electrons sitting in different QDs is not possible. 
Indeed, through the use of conformal field theory and NRG calculations, Ingersent
{\it et al.} were able to characterize a novel, stable, frustration-induced
non-Fermi-liquid phase for a three-impurity Kondo model.\cite{ingersent1} 
It should be noted that this is not the regime treated 
in the current work. Here, we concentrate in the regime where the inter-dot 
hoppings are of the same order of magnitude as the intra-dot Coulomb repulsion, 
and always larger than the coupling to the leads (the `molecular regime').

Considering this rich theoretical context, it is important for our objectives to be clearly spelled out. They are three-fold: 
First, since it is important from an experimental point of view to analyze 
the charge fluctuations in the QDs as a function of the gate potential, and as 
most of the previous work mentioned above uses the Kondo model to represent the QDs, 
we will model the system using the Anderson impurity model to describe each quantum 
dot. Second, we carefully analyze the conductance {\it vs}.\ gate potential 
results in a regime where the inter-dot couplings are larger than the coupling to the leads, 
{\it i.e.}, in the {\it molecular} regime. Although this regime excludes other interesting 
phases in this system analyzed before,\cite{zitko1b} we believe that the molecular regime 
can be experimentally more accessible and therefore very relevant. Third, we analyze in detail 
the effects created by the introduction of a third electron reservoir, which is 
connected to the `free' QD, {\it i.e.}, the QD which is not connected to either of the reservoirs used to measure the 
conductance.

We study this system by calculating the appropriate
propagators to obtain the charge, the local density of states at the dots, and the
conductance, using two different approaches: a finite-U slave boson formalism
developed in the mean-field approximation (FUSBMF),\cite{kotliar} and
the Embedded Cluster Approximation (ECA),\cite{ferrari99,davidovich}
where one diagonalizes a small cluster containing the dots, and then embeds it into
the leads through a Dyson equation. These two completely different
approaches provide a similar description of the physics of the
TQD structure. Note that some of the results shown here were obtained using a
recent variant of the ECA method, the Logarithmic-Discretization 
Embedded Cluster Approximation (LDECA).\cite{number3} In this variant, 
the non-interacting electron band is discretized logarithmically ({\it a la} NRG), which leads to much faster convergence with cluster size. The band discretization provided by LDECA is necessary in two circumstances: i) When finite-size effects preclude ECA from converging to the correct ground state, or when that convergence is too slow; and ii) When one wants to calculate quantitatively accurate local density of states (LDOS). We will clearly indicate when either method (ECA or LDECA) is used.

This paper is organized as follows: In section II, we specify 
the model used to represent the TQD and we briefly describe the numerical 
methods used (FUSBMF, ECA, and LDECA). The results obtained for the case where the 
TQD is coupled to two reservoirs are discussed in section III. The change in the transport properties 
caused by the introduction of a third lead attached to the TQD is discussed in section IV. 
Finally, in section V, we present the conclusions. 

\section{TQD model and numerical methods}

\begin{figure}
\centerline{\resizebox{3.45in}{!}{
\includegraphics{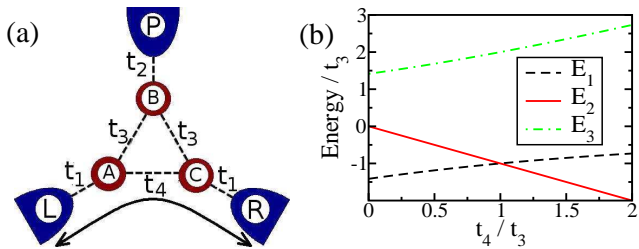}}}
\caption{\label{fig1} (color online) (a) Schematic representation of triple quantum 
dot system coupled to leads. As indicated by the arrows, we will calculate the conductance 
between the left (L) and right (R) contacts. 
(b) Energy diagram (in units of $t_3$) for the molecular orbitals [as defined in Eqs. (5)] as function of 
$t_4/t_3$.}
\end{figure}

The TQD system studied in this work 
is schematically represented in Fig.~\ref{fig1}(a).
The full Hamiltonian can be written as
\begin{eqnarray}\label{hamiltonian}
H=H_{QDs}+H_{leads}+H_{QDs-leads},
\end{eqnarray}
where $H_{QDs}$ describes the isolated TQD system, $H_{leads}$
the two (or three) independent leads and $H_{QDs-leads}$ establishes the contacts between
the dots and the leads. Explicitly, we have
\begin{eqnarray}\label{h_local}
H_{QDs}&=&\sum_{i=A,B,C\atop
\sigma}V_{gi}d^\dagger_{i\sigma}d_{i\sigma}+U\sum_{i=A,B,C}n_{i\uparrow}n_{
i\downarrow}\nonumber\\
&&+\sum_{\sigma}\left[t_3(d^\dagger_{A\sigma}d_{B\sigma}+d^\dagger_{B\sigma}d_{C\sigma})
+t_4d^\dagger_{A\sigma}d_{C\sigma}\right.\nonumber\\
&&\left.+~{\rm H.c.}\right],
\end{eqnarray}
where $d^\dagger_{i\sigma}$ ($d_{i\sigma}$) creates (annihilates) an
electron with spin $\sigma$ in the $i^{th}$ QD and energy controlled by the gate potential $V_{gi}$, $n_{i\sigma}=d^\dagger_{i\sigma}d_{i\sigma}$ is the
occupation number operator, and $U$ is the Coulomb repulsion energy for
double occupancy in a QD.\@
The leads, modeled as semi-infinite chains, are represented by the Hamiltonian,
\begin{eqnarray}\label{h_leads}
H_{leads}=t\sum_{i=1 \atop {\gamma=L,R,P\atop\sigma}}^\infty\left[
c^\dagger_{i_{\gamma}\sigma}c_{(i_{\gamma}+1)\sigma}+~{\rm H.c.}\right],
\end{eqnarray}
where $c^\dagger_{i_{\gamma}\sigma}$ ($c_{i_{\gamma}\sigma}$) creates
(annihilates) an electron with spin $\sigma$ in the $i^{th}$ site of
the $\gamma^{th}$ lead, and $t$ is the kinetic hopping between first
neighbor sites. Finally, the contacts between the QDs and the leads are
established by the Hamiltonian
\begin{eqnarray}\label{h_QDs-leads}
H_{QDs-leads}&=&\sum_{\sigma}\left[t_{1}(
d^\dagger_{A\sigma}c_{1_L\sigma}+
d^\dagger_{C\sigma}c_{1_R\sigma})\right. \nonumber\\
&&\left.+~t_{2}
d^\dagger_{B\sigma}c_{1_P\sigma}+~{\rm H.c.}\right],
\end{eqnarray}
where $c_{1_\gamma\sigma}$ ($\gamma=L,R,P$) annihilates an electron in the
first site of the $\gamma^{th}$ lead. Note also that for all cases studied here, a left $\leftrightarrow$ right symmetry applies.

In order to gain some intuitive understanding of the TQD system, let us consider the non-interacting case first. Let us further assume that $t_1=t_2=0$, so that the QDs are completely
disconnected from the current leads. This is the `atomic limit' of the model and can be solved exactly.
In that case, the Hamiltonian eigenvalues are ($V_{gi}=V_g$),
\begin{subequations}
\begin{eqnarray}
E_1&=&V_g+\frac{t_4}{2}-\frac{1}{2}\sqrt{t^2_4+8t^2_3} \\
E_2&=&V_g-t_4 \\
E_3&=&V_g+\frac{t_4}{2}+\frac{1}{2}\sqrt{t^2_4+8t^2_3}~.
\end{eqnarray}
\end{subequations}
The un-normalized states are given by ([A,B,C]),
\begin{subequations}
\begin{eqnarray}
\mid\psi_1\rangle &=& [1,-\frac{t_4}{2t_3}-\frac{1}{2t_3}\sqrt{t_4^2+8t_3^2},1] \\
\mid\psi_2\rangle &=& [-1,0,1] \\
\mid\psi_3\rangle &=& [1,-\frac{t_4}{2t_3}+\frac{1}{2t_3}\sqrt{t_4^2+8t_3^2},1].
\end{eqnarray}
\end{subequations}

Borrowing the terminology from 
molecular physics, orbitals $\mid\psi_1\rangle$, $\mid\psi_2\rangle$, and $\mid\psi_3\rangle$ 
will be denoted as bonding, non-bonding, and anti-bonding, from now on. 
For the particular case of $t_4=t_3\equiv t^{\prime\prime}$,
$E_1=E_2=V_g-t^{\prime\prime}$ and $E_3=V_g+2t^{\prime\prime}$, the
system has a doubly-degenerate state. In this case, the eigenvalues $E_1$, $E_2$,
and $E_3$ correspond respectively to the orbitals
{\begin{subequations}
\begin{eqnarray}
\mid\psi_1\rangle &=&\frac{1}{\sqrt{6}}[1,-2,1],\\
\mid\psi_2\rangle &=&\frac{1}{\sqrt{2}}[-1,0,1],\\
\mid\psi_3\rangle &=&\frac{1}{\sqrt{3}}[1,1,1].
\end{eqnarray}
\end{subequations}}
The degeneracy results from the symmetry of the system. In group
theory language, it is associated to a two-dimensional irreducible
representation of the $C_{3v}$ symmetry group. Note that, obviously, 
each orbital is also SU(2) symmetric, therefore, at zero field they 
are doubly degenerate regarding the spin orientation. 
For the full interacting Hamiltonian, these orbitals hybridize with
the conduction electron band, renormalizing the eigenvalues. Although
this is a simplified picture, it helps to understand the transport
properties of the system. We will show that the degenerate orbitals have
an important influence in the conductance of the interacting case.

In the interacting case, we are mainly interested in the Kondo regime and will 
study in detail how the symmetry (with its associated degeneracy) affects the transport properties.
As mentioned above, the system is analyzed applying the FUSBMF, 
ECA, and LDECA methods.
All three methods allow the calculation of the Green's functions.
We can easily calculate the charge at each dot 
and the total conductance of the system, which are respectively given by
\begin{eqnarray}
\langle n_{i\sigma}\rangle=\frac{-1}{\pi}\int_{-\infty}^\infty {\rm Im}
\{G^\sigma_{ii}(\omega)\} f(\omega) d\omega
\label{charge}
\end{eqnarray}
 and
\begin{eqnarray}
{\rm G_T}=4\pi^2t_1^4\rho_{R}(\epsilon_F)\rho_{L}(\epsilon_F)\mid
G_{LR}(\epsilon_F)\mid^2,
\label{conductance}
\end{eqnarray} 
 where $G^\sigma_{ii}$ is the local Green's function 
of the QDs, $f(\omega)$ is the Fermi function, $\rho_{L(R)}(\epsilon_F)$
is the density of states of the left (right) lead's first site
and $G_{LR}(\epsilon_F)$ is the Green's function that propagates an 
electron from the left to the right lead, all calculated at the Fermi energy $\epsilon_F$. The
expression for the conductance can be derived from the Keldysh
formalism\cite{wingreen} and is equivalent to the Landauer-B\"uttiker formula for the 
non-interacting case.

\subsection{Finite-U slave bosons mean-field approximation}
In the FUSBMF approach,\cite{kotliar} one enlarges the Hilbert space
by introducing a set of slave boson operators ${\hat e_i}$, $\hat
p_{i\sigma}$ and $\hat d_i$ ($i=A,B,C$), and replace the creation ($
d^\dagger_{i\sigma}$) and annihilation ($ d_{i\sigma}$) operators in
the Hamiltonian by $d^\dagger_{i\sigma}\hat z^\dagger_{i\sigma}$ and
$\hat z_{i\sigma}d_{i\sigma}$, respectively. Following Kotliar and
Rukenstein, the operator $\hat z$ takes the form\cite{kotliar}
\begin{eqnarray}
\hat z_{i\sigma}=[1-\hat d^\dagger_i\hat d_i-\hat p^\dagger_{i\sigma} \hat p_{i\sigma}]^{1/2}
[\hat e^\dagger_i\hat p_{i\sigma}+\hat p^\dagger_{i\bar\sigma}\hat d_i]\nonumber\\
\times[
1-\hat e^\dagger_i\hat e_i-\hat p^\dagger_{i\bar\sigma}\hat p_{i\bar\sigma}]^{1/2}.
\end{eqnarray}
Notice that the bosonic operators $\hat d_i$ and $\hat
e_i$ do not carry spin index.
The enlarged
Hilbert space is then restricted to the physically meaningful subspace by imposing the
constraints
\begin{eqnarray}\label{cP}
\hat P_i=\hat e^\dagger_i \hat e_i+\sum_\sigma
\hat p^\dagger_{i\sigma}\hat p_{i\sigma}+\hat d^\dagger_i\hat d_i-1=0
\end{eqnarray}
 and
\begin{eqnarray}\label{cQ}
\hat Q_{i\sigma}=d^\dagger_{i\sigma}d_{i\sigma}-\hat p^\dagger_{i\sigma}\hat p_{i\sigma}
-\hat d^\dagger_i\hat d_i=0.
\end{eqnarray}
These constraints are enforced by introducing them into
the Hamiltonian through Lagrange multipliers $\lambda^{(1)}_i$ and
$\lambda^{(2)}_{i\sigma}$.  The constraints (\ref{cP}) force the
dots to have empty, single or double occupancy only, and
(\ref{cQ}) relates the boson to the fermion occupancy. In the mean-field 
approximation, the boson operators $\hat e_i$, $\hat
p_{i\sigma}$ and $\hat d_i$ (and the corresponding Hermitian
conjugates) are replaced by their thermodynamical expectation values 
$e_i\equiv\langle\hat e_i\rangle= \langle\hat e^\dagger_i\rangle$,
 $p_{i\sigma}\equiv\langle\hat p_{i\sigma}\rangle=
 \langle\hat p^\dagger_{i\sigma}\rangle$
and  $d_i\equiv\langle\hat d_i\rangle=\langle\hat
d^\dagger_i\rangle$. These expectation values, plus the Lagrange 
multipliers, constitute a set of parameters to be determined by
minimizing the total energy $\langle H \rangle$. In principle, a set of seven selfconsistent parameters are needed for each dot.
Although our system has three dots (which would require a total of
21 parameters), we take advantage of the symmetry of the
configuration, since QDs $A$ and $C$ are symmetric with respect to $B$.
Note that this symmetry imposes no additional constraint on the parameter values. 
In contrast with a previous implementation of this method, \cite{lobos}
our approach allows for a more complete and versatile description of the system 
in terms of its structural parameters. In particular, it can describe the 
transition from non-symmetrical to highly symmetrical regimes as the 
interdot parameters are changed. 

In the mean-field approximation, we can obtain selfconsistent expressions for  the Green's functions.
Then, using Eq.\ (\ref{charge}) and (\ref{conductance}), we can calculate 
the charge at each dot and the conductance.  We note that 
all the calculations with FUSBMF are implemented on the {\it individual} QD basis, while ECA and LDECA utilize the {\it molecular} basis. 

\subsection{Embedded Cluster Approximation}

The ECA method\cite{ferrari99,davidovich} relies on the numerical determination of the ground-state of a cluster with open
boundary conditions. In the following, we briefly sketch details of the method. 

The ECA method tackles  the impurity problem in three steps.
First, the infinite system is naturally cut into two parts: one part {\bf C} (the cluster) contains the interacting region plus as many noninteracting sites of the leads as possible, and a second
part {\bf R} (the `rest'), consisting of semi-infinite chains positioned at left and right in 
relation to the cluster {\bf C}. The number of sites in {\bf C} is denoted by $N_{\mathrm{ED}}$.
Second, Green's functions for both parts are computed independently: current implementations 
of ECA utilize the Lanczos method\cite{elbio} to calculate the interacting Green's function of the interacting region,
while those of the part {\bf R}, being noninteracting, can be computed exactly as well.
In a final step, the artificially disconnected parts are reconnected by means of a 
Dyson equation, which dresses the interacting region's Green's function. 
This step, the actual embedding, is crucial for capturing the many-body physics associated with 
the Kondo effect. Moreover, although the clusters that can be solved 
exactly by means of a Lanczos routine are rather small, 
being of the order of $N_{\mathrm{ED}}\approx 12$ sites only, 
the embedding step successfully compensates for that by dressing the cluster Green's function and 
effectively extending the many-body correlations, induced by the 
presence of the impurity, into the semi-infinite chains {\bf R}. Obviously, strongly correlated 
regimes which depend on extremely low energy scales will be difficult to treat with the 
embedding procedure, although, as mentioned above, great progress has been made lately in this 
respect by introducing a logarithmic discretization procedure into the algorithm.\cite{number3} 
 
We now provide further detail on these steps.
 The Hamiltonians of the left and right semi-infinite, tight-binding chains, {\it i.e.}, the noninteracting {\bf R} part, are described by 
\begin{eqnarray}
H_{\rm sc-L}&=& -t\sum_{l=0,\sigma}^{-\infty} (c_{l\sigma}^\dagger c_{l-1\sigma} +~{\rm H.c.}); \nonumber\\
H_{\rm sc-R}&=& -t\sum_{l=N_{\mathrm{ED}}+1,\sigma}^{\infty}\-\-\- (c_{l\sigma}^\dagger c_{l+1\sigma} +~{\rm H.c.}),
\label{sc}
\end{eqnarray}
where in this notation, the sites  labeled by $i=1,\dots, N_{\mathrm{ED}}$ are inside the cluster {\bf C}.
The semi-infinite chains are connected to the  cluster by the 
following term:
\begin{eqnarray}
H_{\rm hy} &=& -V [c_{1\sigma}^\dagger c_{0\sigma} + c_{N_{\mathrm{ED}}\sigma}^\dagger c_{N_{\mathrm{ED}}+1\sigma}] +~{\rm H.c.},
\label{con}
\end{eqnarray}
where $V=t$ is the hopping in the broken link, connecting parts {\bf R} and {\bf C}, 
used in the embedding procedure.  The Green's function for 
the  cluster {\bf C} and for the semi-infinite chains are calculated at zero temperature. 
Fixing the number of particles $m$ and the $z$-axis projection of the total spin, $S_{\mathrm{total}}^z$, the ground
state and the one-body propagators between all the clusters' sites are calculated.  
For example, $g_{ij}^{(m,S^z_{total})}$, the undressed Green's function for the cluster, propagates a particle 
between sites $i$ and $j$ inside the cluster. 
For the noninteracting, semi-infinite chains, 
the Green's functions $g_{0}^L$ and $g_{N_{\mathrm{ED}}+1}^R$ at the sites $0$ and $N_{\mathrm{ED}}+1$, located at the extreme ends 
of the semi-infinite chains, at left and right to the cluster, can be easily calculated as well.

The Dyson equation to calculate the dressed Green's function matrix elements $G^{(m,S_{\mathrm{total}}^z)}_{i,j}$ 
can therefore be written as 
\begin{eqnarray}
G^{(m,S_{\mathrm{total}}^z)}_{i,j} &=& g^{(m,S_{\mathrm{total}}^z)}_{i,j} +
g^{(m,S_{\mathrm{total}}^z)}_{i,1}~V~G^{(m,S_{\mathrm{total}}^z)}_{0,j} \nonumber  \\
   && +
g^{(m,S_{\mathrm{total}}^z)}_{i,N_{\mathrm{ED}}}~V~G^{(m,S_{\mathrm{total}}^z)}_{N_{\mathrm{ED}}+1,j}
\label{Dyson1}
\end{eqnarray}
\begin{subequations}
\label{Dyson2}
\begin{eqnarray}
G^{(m,S_{\mathrm{total}}^z)}_{0,j} &=& g_0^L~V~G^{(m,S_{\mathrm{total}}^z)}_{1,j} \\
G^{(m,S_{\mathrm{total}}^z)}_{N_{\mathrm{ED}+1},j} &=& g_{N_{\mathrm{ED}}+1}^R~V~G^{(m,S_{\mathrm{total}}^z)}_{N_{\mathrm{ED}},j}~,
\end{eqnarray}
\end{subequations}
where $V$, as mentioned above, is defined according to $H_{\rm hy}$. 
Eqs.~(\ref{Dyson1}) and (\ref{Dyson2}) correspond to a chain approximation, where a locator-propagator
diagrammatic expansion is used.\cite{anda81,metzner91} Note that ECA is exact in the case of $U=0$.

As mentioned before, the calculation of the propagator $g^{(m,S_{\mathrm{total}}^z)}_{i,j}$ 
requires that fixed quantum numbers $m$ and $S_{\mathrm{total}}^z$ be used. However, after
the embedding procedure, these quantum numbers are not good quantum numbers for the cluster anymore. 
Therefore, we have to incorporate 
processes into the ECA method  that allow for  charge fluctuations in the cluster {\bf C}.
To accommodate this requirement, different implementations of ECA have been devised, either by including  
different spin mixing strategies\cite{ferrari99,Fabian} or by moving the Fermi energy of the leads.\cite{Willy}

The spin mixing proceeds as follows. First, a cluster Green's function with mixed charge is defined through 
\begin{equation}
g^{(m+p,pS_{\mathrm{total}}^z)}_{i,j} = (1-p)~g^{(m,0)}_{i,j} ~+~ p~g^{(m+1,S_{\mathrm{total}}^z)}_{i,j},
\label{p-equat}
\end{equation}
where $p$ takes values between 0 and 1, and we are assuming that $m$ is even, in which case, 
the corresponding $S_{\mathrm{total}}^z=0$. In addition, note that 
for the cluster with charge $m+1$, $S_{\mathrm{total}}^z$ takes values $\pm 1/2$. 
The matrix element $g^{(m+p,pS_{\mathrm{total}}^z)}_{i,j}$ corresponds to a situation where 
the charge in the cluster is between
$m$ and $m+1$. The total charge in the cluster, before embedding, can be easily calculated as
\begin{equation}
q^{pS_{\mathrm{total}}^z}(p) = (1-p)m ~+~ p(m+1) = m+p.
\end{equation}
Using Eqs.~(\ref{Dyson1}) and (\ref{Dyson2}), the dressed
Green's function $\hat{G}^{(m+p,pS_{\mathrm{total}}^z)}_T$ is obtained, and from this result, 
the charge in the cluster can be calculated:
\begin{equation}
Q^{pS_{\mathrm{total}}^z}(p) = \frac{-1}{\pi} \int_{-\infty}^{E_F} 
\mbox{Im}~\{\sum_{i}G^{(m+p,pS_{\mathrm{total}}^z)}_{i,i}(\omega)\} d\omega\,,
\end{equation}
where $E_F$ is the Fermi level. The value of $p$ is calculated self-consistently, satisfying
\begin{equation}
Q^{pS_{\mathrm{total}}^z}(p) = q^{pS_{\mathrm{total}}^z}(p).
\label{selfcons}
\end{equation}
If there is spin reversal symmetry, e.g., no magnetic field is applied, one can calculate the total Green's function as
\begin{equation}
G_{i,j}^T(p) = {1 \over 2} \sum_{S_{\mathrm{total}}^z=\pm 1/2} G^{(m+p,pS_{\mathrm{total}}^z)}_{i,j}, 
\end{equation}
where $p$ satisfies Eq.~(\ref{selfcons}).
It is important to emphasize that the charge fluctuations taken into account by Eq.~(\ref{p-equat})
are the ones between the cluster and the rest of the system, and not just the ones at the interacting
region described by $H_{\rm int}$. The latter ones involve a very localized neighborhood of the dot and as a consequence,
 are typically already well described on isolated clusters only.
Finally, it is noteworthy to point out that the self-consistent solution for the charge mixing parameter $p$ 
is either 0 or 1 in the Kondo regime, and in particular at the particle-hole symmetric point $V_g=-U/2$ 
(when analyzing a single-QD problem). 
Therefore, deep into the Kondo regime ({\it i.e.}, at the particle-hole symmetric point), no charge mixing takes place at all, 
and very little charge mixing occurs in a window of gate potential around the particle-hole symmetric point. 
The parameter $p$ will start to take a finite value (note that $0 \leq p \leq 1$) as the gate potential 
drives the system into the mixed-valence regime.  The purpose of the 
charge mixing is thus mainly to smooth out the transition from 
an $N$ electron to an $N\pm 1$ electron ground state, which for the bare
cluster is a crossover between ground states with different number of particles.

As mentioned above, some of the calculations 
were done using the LDECA method, which is an important extension of ECA. 
In it, to obtain a better description of the low energy physics of the system, 
the non-interacting band is logarithmically discretized. All the procedure 
described above remains the same, but the band discretization allows a 
much faster convergence to the Kondo regime with cluster size. A full 
description of LDECA can be found in Ref.~\onlinecite{number3}. 

\begin{figure}
 \begin{center}
\resizebox{3.5in}{!}{\includegraphics{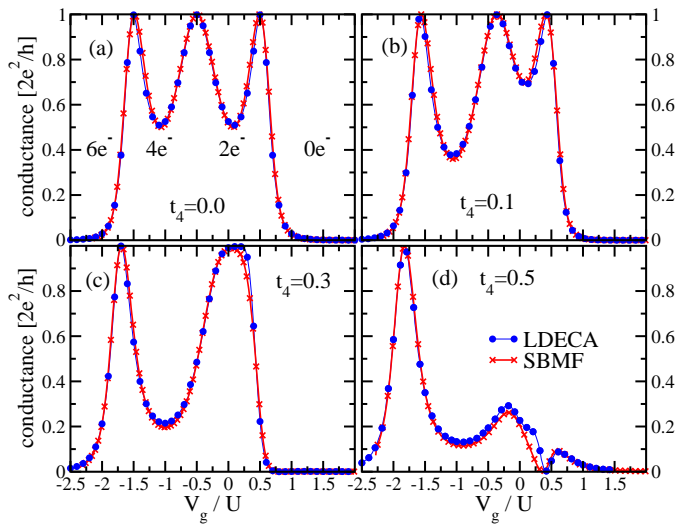}}
\end{center}
\caption{(color online) Conductance as function of gate potential,
$V_g=V_{gA}=V_{gB}=V_{gC}$, for various values of $t_4$, as indicated 
in the panels. (Blue) solid dots indicate LDECA results for clusters with 11 sites; 
(red) `$\times$' signs are FUSBMF results. Notice in panel (a), where $t_4=0$, 
that the number of electrons in the TQD system is indicated for each 
one of the valleys, where the Kondo effect is absent.
The structure evolves from three to two peaks
as the system goes from three dots in series [$t_4=0.0$, panel (a)], to isosceles [$t_4 < t_3=0.5$, panels (b) and (c)], 
and finally to equilateral triangle symmetry [$t_4=t_3$, panel (d)]. For all panels, 
$U=1.0$, $t_1=0.45$, $t_2=0.0$, and $t_3=0.5$. 
The very small discrepancy between LDECA and FUSBMF in panel (d) comes 
from a finite-size effect in the LDECA results. }
\label{fig2}
\end{figure}

\subsection{Numerical results}
In order to study the conductance as a function of the parameters of
the system, we use the leads' hopping coupling $t$ as the energy unit 
($t=1$), and set the Fermi energy to zero ($\epsilon_F=0$). All the
results will be shown for zero-temperature. In the strong interdot coupling regime, 
{\it i.e.}, $t_3,t_4 > t_1,t_2$, the individual QDs levels mix into three molecular
orbitals which are coupled to the leads. The energy of these
orbitals can be controlled by gating the local energy states and by varying the
hopping matrix elements ratio $t_4/t_3$ [see Fig.~\ref{fig1}(b)]. Their widths
depend upon the coupling to the conduction bands, {\it i.e.}, $t_1$ and $t_2$. In
order to study the contribution of each individual orbital to the conductance, 
we make them sufficiently far apart from each other. To do so, we
take $t_3>t_1,t_2$. In particular, in section III, we set $t_1=0.45$, $t_2=0.0$, $t_3=0.5$, 
and $U=1.0$ (which are the same parameters used in Ref.~\onlinecite{lobos}), 
and in section IV we set $t_1=0.2$, $0.0 \leq t_2 \leq 0.2$, $t_3=0.4$, and $U=0.5$.\cite{note-molecular} 
In section III, we vary $t_4$ to manipulate the symmetry of the system ($0.0 \leq t_4/t_3 \leq 1.0$). 
In section IV, besides the same variation of $t_4$ as in section III, we also analyze what is the 
effect of varying $t_2$ ($0.0 \leq t_2/t_1 \leq 1.0$), {\it i.e.}, we verify what is the 
effect of adding a third lead [connected to QD B, see Fig.~\ref{fig1}(a)] to the TQD system. 

\section{TQD connected to two leads ($t_2=0$)}

\subsection{TQD in series ($t_4=0$)}

Initially, taking advantage of the flexibility of the numerical methods used,
we analyze the conductance when the three QDs are aligned in series ($t_4=0$) 
and coupled to two leads only ({\it i.e.}, we make $t_2=0$). Fig.~\ref{fig2} shows the conductance
as a function of the gate potential $V_g=V_{gA}=V_{gB}=V_{gC}$. In panel (a), for
$t_4=0$ [(red) `$\times$' signs indicate FUSBMF results, and (blue) solid dots display LDECA results], 
the three molecular orbitals are equally separated by an energy value proportional to $t_3$.
As will be shown below, this originates from three Kondo peaks (occurring at different $V_g$ values) 
associated to each molecular orbital. 

\begin{figure}
\includegraphics[width=3.5in]{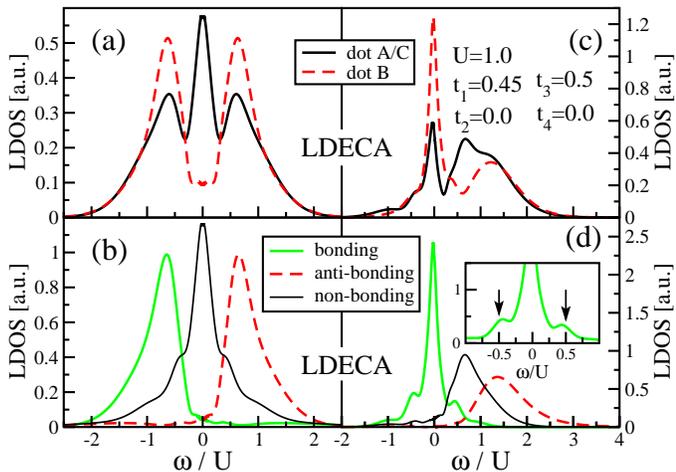}
\caption{(color online) LDOS results using LDECA (9 sites) for $t_1=0.45$, $t_2=0.0$, $t_3=0.5$, $t_4=0.0$, and $U=1.0$. 
(a) and (c): LDOS for each QD. (a) and (b) are calculated at $V_g=-U/2$, while (c) and (d) at $V_g=U/2$. 
Note that, in (a), QDs A and C are in Kondo and QD B is not, while in (c) all 
dots participate in the Kondo effect. 
(b) and (d): LDOS for each molecular orbital. In (b), just the non-bonding orbital is in the Kondo 
state (dark (black) curve), as one can easily see its Kondo peak at $\omega=0.0$, while 
the bonding orbital (light (green) curve) is almost fully occupied (most of its LDOS is below the 
Fermi energy) and the anti-bonding orbital (dashed (red) curve) is almost empty (most of its 
LDOS is above the Fermi energy). In (d), it is now the bonding state that has 
a Kondo resonance, while the non-bonding and anti-bonding states are higher in energy. 
The inset shows a more detailed view of the Kondo peak for the bonding orbital, 
where the small LDOS peaks for the $\epsilon$ and $\epsilon+U$ states are indicated by arrows.
}
\label{fig3}
\end{figure}

The  $t_4=0$ case corresponds to the molecular regime reported in
Ref.~\onlinecite{zitko2} (with, as mentioned above, $t_3>t_1$). 
For $t_4\neq0$, the system transforms into a
triangular configuration, which can be compared to the system
studied in Ref.~\onlinecite{lobos}. In the first case 
($t_4=0$, panel (a) in Fig.~\ref{fig2}), the three
peaks in the conductance occur at gate potential values 
where there is a change in the occupation of the different molecular 
orbitals. When the bonding orbital hosts one electron
(for $V_g/U=0.5$), the system is in the Kondo regime and the
characteristic Abrikosov-Suhl resonance of this regime creates a
path for the electrons to cross from the left (L) to the right (R) lead 
[this will be more clearly demonstrated in Fig.~\ref{fig3}(d)]. 
Decreasing $V_g$, we find a valley corresponding to the
accommodation of a second electron in the bonding state, creating a
singlet that destroys the Kondo effect. The middle peak corresponds
to the presence of a third electron in the system, now siting in the
non-bonding state, since the bonding state is full. Again, the
conductance peak reflects the Kondo resonance at the Fermi level due
to an unpaired electron that is anti-ferromagnetically correlated with
the conduction electrons (see Fig.~\ref{fig3}(b), and discussion 
below). The following valley and the third peak
are a consequence of the suppression of the Kondo effect due to
double occupation of the non-bonding  orbital and the unpaired
electron in the anti-bonding orbital, respectively. Finally, the final drop
in the conductance results from the destruction of the Kondo effect
due to the sixth electron entering into the system. The electron occupancies 
at the conductance valleys are indicated in panel (a).

\subsection{LDECA LDOS results for molecular orbitals ($t_4=0$)}

Before analyzing the effect of introducing a finite $t_4$, we want to show 
LDOS results at the particle-hole symmetric point ($V_g=-U/2$) [Figs.~\ref{fig3}(a) and (b)], and at 
$V_g=U/2$ [Figs.~\ref{fig3}(c) and (d)] for the 
$t_4=0.0$ curve in Fig.~\ref{fig2}(a). In the upper left panel in Fig.~\ref{fig3}, we have the LDOS 
for each QD for $V_g=-U/2$. As expected, QDs A and C have the same LDOS (dark (black) solid curve), 
and the peak at $\omega=0$ is indicative 
that they participate in a Kondo effect. Indeed, a clear Kondo peak 
can be seen at the Fermi energy $E_F$ ($\omega=0.0$) for QDs A and C, while the LDOS for QD B 
has a gap at $E_F$ [dashed (red) curve in panel (a)]. This will be important later on to understand the 
results when the upper lead (P) is coupled to QD B (for finite $t_2$). 
The LDOS for the appropriate 
orbital states for this configuration ($t_4=0.0$, and $V_g=-U/2$) is shown in the lower left panel. The bonding orbital (gray (green) solid curve) 
has most of its LDOS below $E_F$, indicating that it is already almost fully occupied. In contrast, 
the anti-bonding orbital has, at this particular gate potential, most of its LDOS above 
$E_F$ and is therefore almost completely empty. The non-bonding orbital [which is an antisymmetrical 
combination of QDs A and C, only, see Eq. 6(b)] displays a Kondo peak at $E_F$, which is responsible 
for the unitary conductance seen for $t_4=0$ in Fig.~\ref{fig2}(a) at $V_g=-U/2$. 
As to the rightmost peak in Fig.\ \ref{fig2}(a), notice, as can be seen in the upper right panel 
in Fig.\ \ref{fig3}(c), that all three QDs participate in the Kondo effect for $V_g = U/2$. 
In the lower right panel, one sees the LDOS for the orbital states, 
now indicating that the bonding state has a Kondo peak, while the other two 
orbitals are nearly empty.\cite{edge} The inset shows details of the Kondo peak, for the bonding orbital, including the 
shorter peaks associated to $\epsilon$ and $\epsilon+U$, indicated by vertical arrows. 

\subsection{Finite $t_4$ and interference effects ($t_4 \approx t_3$)}

Now, as $t_4$ increases from $0$ to $0.5$ (panels (b) to (d) in Fig.~\ref{fig2}), the bonding and non-bonding
energies become closer to each other [see Eqs. (5)], and finally, become degenerate
for $t_4=t_3=0.5$ [see Fig.~1(b)], when the system possesses an equilateral triangle
symmetry. Note that the peaks in the conductance curves 
shown in Fig.~\ref{fig2}, corresponding to
the bonding (rightmost peak) and non-bonding (central peak) orbitals, merge into each other and the conductance
decreases as $t_4\rightarrow t_3$. 
As shown above in Fig.~\ref{fig3}, the LDOS of the molecular orbitals provides a 
more clear picture of the Kondo effect than the LDOS of each QD. For example, when $t_4=0$, 
the rightmost conductance peak in Fig.~\ref{fig2}(a) 
can be directly associated to the Kondo peak in Fig.~\ref{fig3}(d) (light (green) solid 
curve). As mentioned above, the molecular orbitals provide a natural description
of the Kondo effect when the intra-dot hoppings are larger than the coupling to the 
leads. 

In the same manner that the LDOS for each molecular orbital provides important 
insight into the conductance through the TQD system, one can define a `partial' 
conductance ${\rm G_i}$ through each molecular orbital `${\rm i}$' (${\rm i}=1,2,3$) in the following way (full details 
are given in Ref.~\onlinecite{interf}): 

\begin{equation}
{\rm G_i} = \frac{e^2}{h} \left[ t^2 \tilde{g}_l t_{li} \rho(\epsilon_F)\right]^2 |G_{iR}(\epsilon_F)|^2 
\label{eq-Gi}
\end{equation}
where $\tilde{g}_l$ is the Green's function in the first site of the left contact, 
$t_{li}$ is the coupling of the left lead with orbital $\mid \psi_i \rangle$, and $G_{iR}$ 
is the dressed Green's function that moves an electron from $\mid \psi_i \rangle$ 
(where $i=1,2,3$) to the first site in the {\em right} contact. For $t_4 \approx t_3$ and $V_g$ values such that 
mostly molecular orbitals $i=1,2$ are involved in the transport of charge ({\it i.e.}, $-0.5 \lesssim V_g/U \lesssim 0.5$, 
in panels (c) and (d) in Fig.~\ref{fig2}), the total conductance ${\rm G_T}$ can be approximated by the equation 
\begin{equation} 
{\rm G_T} \simeq {\rm G}_{12}={\rm G}_1+{\rm G}_2+2\sqrt{{\rm G}_1{\rm G}_2}\cos{\Delta \phi_{12}},
\label{eq-interf} 
\end{equation}
where 
\begin{equation}
i \Delta \phi_{12} = \log \left\{ \frac{G_{1R}}{G_{2R}} \frac{|G_{2R}|}{|G_{1R}|} \right\}
\label{eq-phase}
\end{equation}
defines the phase-difference between a path that goes through orbital $\mid \psi_1 \rangle$ and 
a path that goes through orbital $\mid \psi_2 \rangle$. In the case where all three orbitals are contributing, a 
simple extension of these equations should be used, and it gives results exactly equal to the ones 
shown in Fig.\ \ref{fig2}. 
We should note that the ${\rm G_i}$  functions have a characteristic `width' given by the coupling to the leads (and the weight of the orbital at the connecting dot), as well as a position dependence on $V_g$, as the energy of each orbital shifts with respect to the Fermi energy.

\begin{figure}
\vspace{3mm}
\includegraphics[width=3.4in]{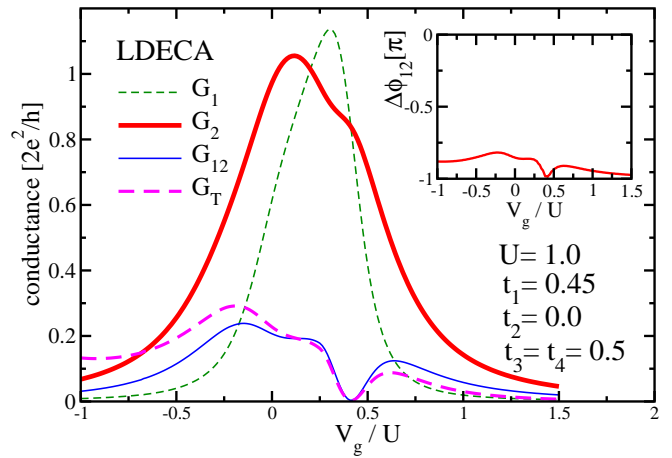}
\caption{(color online) Detail of the conductance (calculated with LDECA -- 11 sites) 
around the degenerate states $\mid \psi_1 \rangle$ and $\mid \psi_2 \rangle$:
${\rm G}_1$ (short dash (green) curve) and ${\rm G}_2$ (thick solid (red) curve), 
as defined in Eq.\ (22), ${\rm G}_{12}$ (thin solid (blue) curve), from Eq.\ (\ref{eq-interf}),
and the total conductance (involving all molecular orbitals) ${\rm G_T}$ (long dash (magenta) curve) from Eq.\ (9). 
The inset shows the phase difference between electrons propagating through 
molecular orbitals $\mid \psi_1 \rangle$ and $\mid \psi_2 \rangle$.  See text for details.}
\label{fig4}
\end{figure}

Equation~(\ref{eq-interf}) shows that when there is no energy overlap between 
$\mid \psi_1 \rangle$ and $\mid \psi_2 \rangle$ (therefore, no 
overlap between ${\rm G}_1$ and ${\rm G}_2$), which occurs when the corresponding 
orbitals are well separated in energy, the last 
interference term is zero, {\it independently} of the value of the phase difference. 
This is essentially the case for the conductance results for $t_4=0$ in Fig.\ \ref{fig2}(a). 
In this case, a simple sum of the partial 
conductances ${\rm G_i}$ through each molecular orbital is very similar (not shown) to the total conductance 
(and more so as the level separation increases for larger values of $t_3/t_1$). However, as shown 
next (see Fig.~\ref{fig4}), once the molecular orbital levels start to overlap, the 
partial conductances ${\rm G_i}$ are no longer simply related to the total conductance, 
as the last term in Eq.~(\ref{eq-interf}) now plays a role, and its effect will 
obviously depend on the value of $\Delta \phi_{12}$. Therefore, the calculation of the partial 
conductances, and the phase difference of the corresponding Green's functions, provides us with 
information about possible interference effects, as shown next. However, a word of caution is 
necessary. Since the simple addition of the partial conductances does not reproduce the total conductance 
when there is overlap between the molecular levels, we will not discuss the details 
of the ${\rm G_i}$'s, as they do not, by themselves, describe an experimentally observable quantity. Obviously, 
when there is no overlap, as is the case for $t_3 \gg t_1$ and $t_4=0$ (see section IV), the partial conductance 
of each orbital is identical to the total conductance.

Figure~\ref{fig4} shows, in the main panel, LDECA partial conductances 
${\rm G}_1$ (short-dashed (green) line) and ${\rm G}_2$ (thick solid (red) line), for molecular orbitals $\mid\psi_1\rangle$ and 
$\mid\psi_2\rangle$, respectively, the {\it total} conductance ${\rm G_T}$ (long-dashed (magenta) line), which 
takes in account all 3 orbitals, and ${\rm G}_{12}$ (thin solid (blue) line), as obtained through 
Eq.~(\ref{eq-interf}), where just orbitals $\mid\psi_1\rangle$ and $\mid\psi_2\rangle$ are taken in account. 
The reason why ${\rm G}_{12}$ and ${\rm G_T}$ are so similar 
is because orbital $\mid\psi_3\rangle$ is at a considerably higher energy 
in relation to the degenerate orbitals and therefore its contribution to 
the conductance for gate potential values around $V_g = 0.5U$ is
minimal. The phase difference $\Delta \phi_{12}$ (in units of $\pi$), as a function of gate 
potential, is shown in the inset. 
Note that the dip in the total conductance is related to a $V_g$ value where both 
partial conductances have the same value and $\Delta \phi_{12}=\pi$ (see inset).\cite{g0}

The conductance features described in Fig.~\ref{fig2} are quite different
from the results reported by the authors in Ref.~\onlinecite{lobos}.
Although their system is the same as ours (and the parameters are the same), the authors of Ref.~\onlinecite{lobos}
consider, for simplicity, a regime where the total molecular region is described by a single level
impurity, which allowed them to reduce the number of bosons in the FUSBMF approximation.
However, as our results show, the details of the internal structure
of the molecule are essential to determine its transport properties. Notice that
as $t_4$ increases, not only the peaks shift their positions but, as
previously mentioned, also the structure of the peaks changes. 

Panels (c) and (d) in Fig.~\ref{fig2} display slight quantitative discrepancies between LDECA and FUSBMF results, 
although the overall qualitative agreement between the two techniques is quite good. 
These discrepancies stem from finite-size effects in the LDECA results.\cite{finite}
LDECA calculations for increasingly larger exactly diagonalized clusters (not shown) indicate that the LDECA results gradually 
approach those from FUSBMF. This convergence becomes slower as $T_K$ decreases, but the 
LDECA and FUSBMF qualitatively agree for all regimes we checked. 

Note that in the limit of strong coupling between
dots A and C ($t_4 \gg t_3$, and for $U > 2t_4$), a two-stage Kondo regime
(TSK) should be expected (at half-filling).\cite{cornaglia,number3} 
In this regime, dot B is weakly coupled to the
band through the Kondo resonances of quantum dots A
and C, producing a second Kondo stage, with an exponentially smaller
characteristic energy $T_{\rm TSK} \ll T_{\rm K}$. This special regime will be analyzed in a future work.

\subsection{S=1 Kondo effect ($t_4=t_3$)}

{In addition to interference, the degeneracy (caused by symmetry) 
has an additional effect: it causes the two degenerate orbitals 
(when occupied by one electron each) to develop a ferromagnetic correlation. 
\cite{ferro} Indeed, when the structure reaches the equilateral
triangle symmetry ({\it i.e.}, $t_4=t_3=0.5$), the two degenerate molecular orbitals are charged 
simultaneously. In this case, due to the Kondo correlation, the
first two electrons enter in the system with parallel spins, and the
system presents an $S=1$ Kondo effect. 
This can be quantitatively appreciated by calculating the total spin for the three 
QDs as a function of gate potential, as well as the individual occupancy of each of the 
three molecular orbitals. This is shown by the LDECA results in Fig.~\ref{fig5}, where, together 
with the total conductance (long dash (magenta) curve), the charge occupancy for each 
of the molecular orbitals is shown (thin solid (purple) curve for orbital $\mid\psi_1\rangle$, 
dot-dashed (black) curve for orbital $\mid\psi_2\rangle$, and short dash (blue) curve for orbital $\mid\psi_3\rangle$), 
and the total spin in the three QDs (thick solid (green) curve). Notice that the occupancy dependence with $V_g$ 
for the two degenerate orbitals is not identical because they couple differently to 
the leads (orbital $\mid\psi_2\rangle$ couples more strongly than $\mid\psi_1\rangle$). 
As mentioned above, the maximum in the value of the total spin ($S_T \approx 0.7$, see thick solid (green) curve) occurs when 
there is approximately one electron\cite{note-double} in each of the degenerate orbitals, 
which couple through an {\it effective} ferromagnetic 
interaction.\cite{ferro} (Note that a value of $S=1$ 
will not be obtained for such large ratios of hopping over Coulomb repulsion). 
This spin configuration reduces the ground state energy by Kondo 
correlating the total $S=1$ spin with the conduction electrons. In this region 
of gate potential, the system is in the Kondo regime, which provides a 
way for the electrons at the Fermi level to cross from QD A to QD C. However, 
having two interfering channels at their disposal, constructed from the two degenerate 
orbitals, the conductance (long dash (magenta) curve) possesses a very clear 
Fano-like antiresonance. For lower gate potential values ($V_g \approx -1.75$), 
only orbital $\mid\psi_3\rangle$ is involved in electron transport and therefore the conductance 
has the usual Lorentzian shape, with maximum value $G_0$, and $S_T \approx 0.4$. 

\begin{figure}
\includegraphics[width=3.5in]{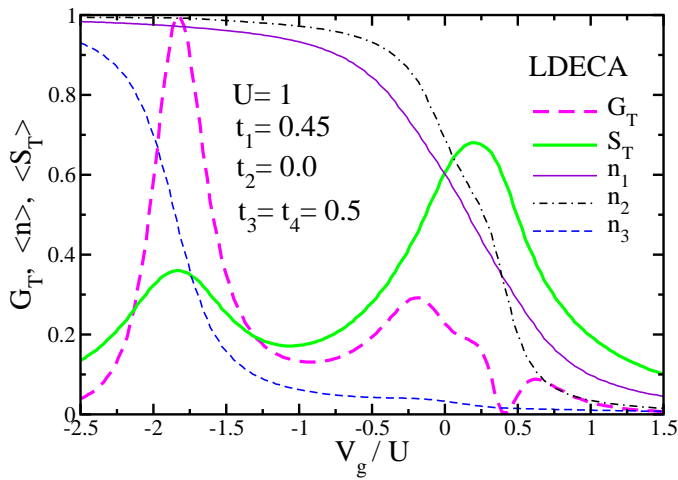}
\caption{(color online) LDECA results (11 sites) for conductance 
(long dash (magenta) curve), same parameters as in Fig.~\ref{fig4}. 
Charge per spin for each orbital [$\mid\psi_1\rangle$, thin solid (purple), 
$\mid\psi_2\rangle$, dot-dash (black), and $\mid\psi_3\rangle$, short dash (blue)], 
and total spin [thick solid (green)], as a function of gate potential. See text for details.}
\label{fig5}
\end{figure}

\subsection{Deeper into Kondo and molecular regimes}

As mentioned above, the parameters in this section were chosen to match 
those in Ref.~\onlinecite{lobos}. As expected, and clearly demonstrated 
by the LDOS's in Fig.~\ref{fig3}, the TQD system for these parameters 
seems to be closer to the intermediate valence regime than to the Kondo 
regime. Also, based on the fact that $t_1 \approx t_3$, one may question 
if the molecular orbitals are really the most appropriate description of 
the single electron properties of the system. In view of that, in section IV, 
where the effect of introducing a third lead will be analyzed, the parameters 
will be changed so that the system will be deeper into the Kondo regime 
(with a larger $U/\Gamma$ than in the current section). To accomplish that, we will choose 
$U=0.5$ and $t_1=0.2$. In addition, in the next section, we will choose 
$t_3=0.4$ (with $0 \leq t_4 \leq t_3$), which brings the system 
more effectively into the molecular regime (as $t_3/t_1=2$). To illustrate both points, 
Fig.\ \ref{fig6} shows the same LDOS results as in Fig.~\ref{fig3}, 
but now for the new parameter set. It is apparent that the Kondo peaks for the new parameters are more well defined. 
For example, compare the dark solid (black) curves in panel (b) of both figures, 
which display the Kondo peak for the non-bonding orbital ($\mid\psi_2 \rangle$). 
The peak in Fig.~\ref{fig6}(b) clearly shows a sharper structure at the Fermi energy 
than the one in Fig.~\ref{fig3}(b), indicating that this system is 
deeper into the Kondo regime. It is also apparent that the LDOS of the different 
molecular orbitals have much less overlap in Fig.~\ref{fig6}, underscoring the 
fact that, for the parameters to be used in section IV, 
the molecular orbitals provide a more suitable description of the TQD system.  
Nonetheless, notice that the molecular orbitals 
provide an appropriate framework to understand the results presented in section III as well, 
as it is clear that Figs.~\ref{fig3} and \ref{fig6} are {\it qualitatively} similar.

\begin{figure}
\includegraphics[width=3.5in]{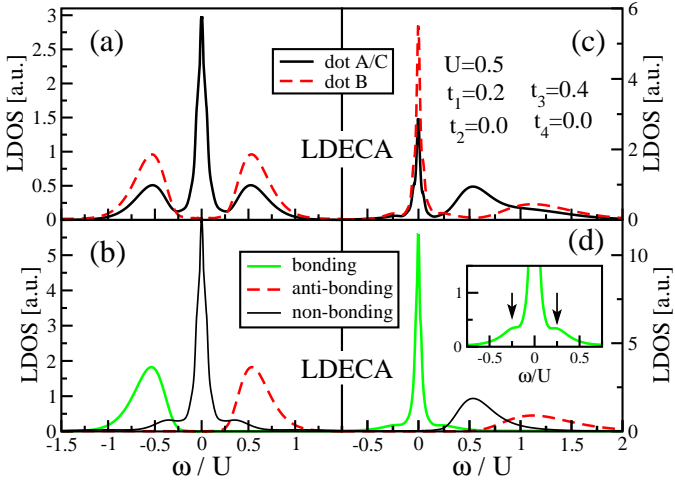}
\caption{(color online) Same LDECA calculations (9 sites) as in Fig.~\ref{fig3}, but now for 
$U=0.5$, $t_1=0.2$, $t_2=0.0$, $t_3=0.4$, and $t_4=0$. Note that the Kondo peaks are better defined than the ones in Fig.\ \ref{fig3}, as there is less overlap of the LDOS from different molecular orbitals, indicating that for these parameters the system is deeper 
into the Kondo and molecular regimes. Nonetheless, the qualitative 
similarities with Fig.~\ref{fig3} are evident, indicating that 
the molecular orbitals are appropriate for the description of the results 
in section III. 
}
\label{fig6}
\end{figure}

{
\section{Loss of amplitude through a third lead}
}

\begin{figure}
\begin{center}
\begin{tabular}{cc}
\resizebox{3.5in}{!}{\includegraphics{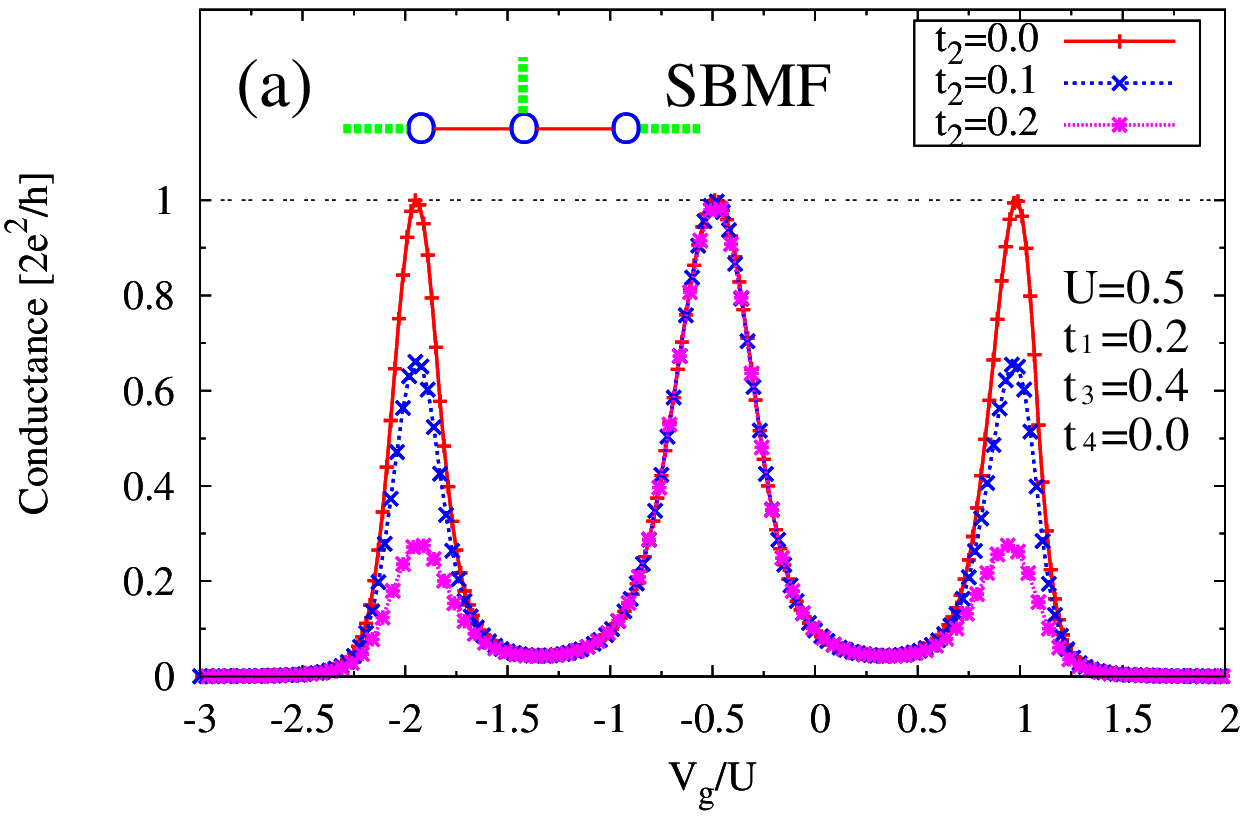}}\\
\resizebox{3.5in}{!}{\includegraphics{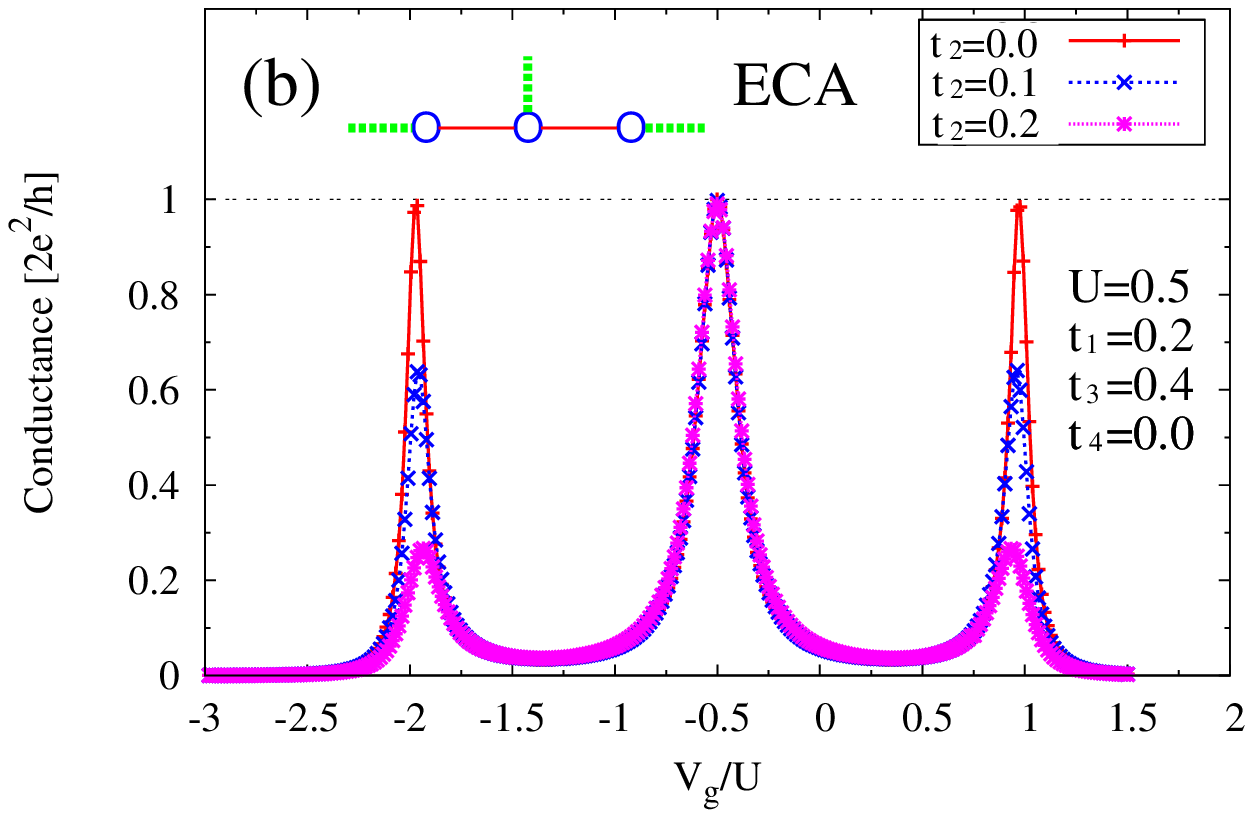}}
\end{tabular}
\end{center}
\caption{\label{fig7} (color online) Conductance as function of gate potential,
$V_g=V_{gA}=V_{gB}=V_{gC}$, for different values of $t_2$, showing the effect of the
`amplitude loss' due to the presence of the third lead. The other 
parameters are $U=0.5$, $t_1=0.2$, $t_3=0.4$, and $t_4=0$. 
Panels (a) and (b) refer to FUSBMF and ECA (6 sites) results, respectively.}
\end{figure}

In this section, as just mentioned, we use different parameters ($t_1=0.2$, $t_3=0.4$, $U=0.5$) from the ones used 
in section III. The objective is to have a larger 
value of $U/\Gamma$, and therefore move deeper into the Kondo regime 
and away from the intermediate valence. 

Based on a comparison of the results in Fig.~\ref{fig2} with those in Figs.~\ref{fig7} and \ref{fig8}, 
a clear picture emerges of the effect of a third lead connected to QD B [see Fig.~\ref{fig1}(a)]. 
Using the labels defined in Fig.\ 1(a) for the QDs, let us 
qualitatively describe how the coherent propagation of electrons is affected 
by the additional lead. Assume that an electron is traveling from the left into QD A.\@ After arriving at QD A, the electronic wave splits 
into two: one travels via QD C, and the other 
via QD B. The latter portion, on reaching 
QD B, will be split into two again: one travels {\it away} through 
the upper lead, while the other travels via QD C. 
We can view this process of `electron loss' through lead P (the `third' lead) 
as being a process of `amplitude leakage', like that occurring at a beam splitter. 
The remaining two traveling waves (traveling through 
the triangle, in the direction of QD C) are coherent and will 
interfere when they propagate out of the system through 
the right lead. It will be shown below that the introduction of lead P does not make 
the electron propagation incoherent, since the propagation 
through overlapping molecular orbital levels clearly shows signs 
of interference, the same way as observed for $t_2=0$,  when the third
lead is absent, as was discussed in Figs.~\ref{fig2} and \ref{fig4}.
To analyze the results for conductance and LDOS, we use again the molecular orbital basis. 
The strategy for this analysis can be summarized by the following two observations:
First, by analyzing the conductance through each molecular orbital, 
one realizes that the percentage of the traveling wave lost through 
lead P will depend on the coupling of each  molecular 
orbital to it. This `loss' through lead P will result in a
lower partial conductance through the molecular orbital in question.
Note that, for a fixed value of $t_2$, the coupling to lead P depends 
only on the coefficient of QD B in each molecular orbital, 
which varies with the ratio $t_4/t_3$.
Second, if we assume that the transport through 
each molecular orbital is coherent (even after coupling 
QD B to lead P), the transport through two overlapping 
molecular orbital levels should give origin to interference 
effects, as in the case where lead P is not present 
(see section III). We will show evidence below that this is indeed 
the case.

\begin{figure} 
\centerline{\resizebox{3.5in}{!}{
\includegraphics{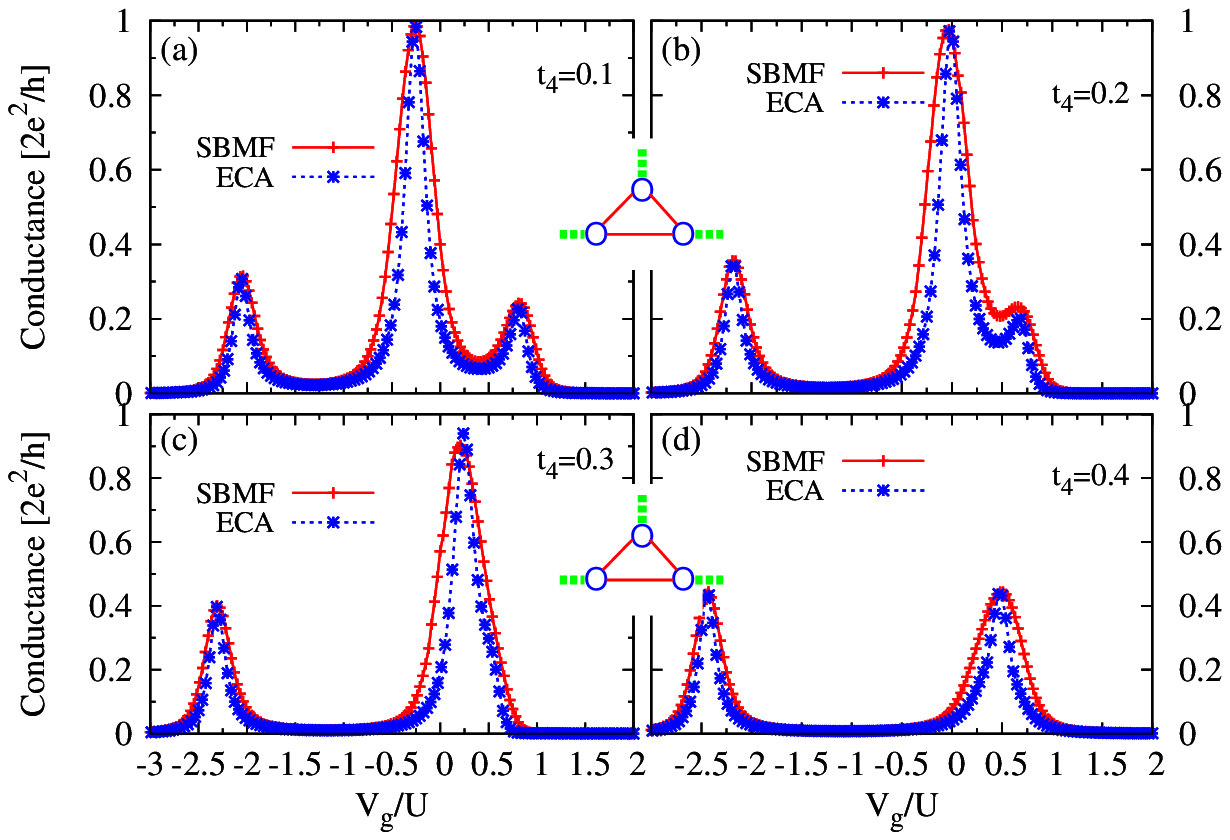}}}
\centerline{\resizebox{3.5in}{!}{
\includegraphics{fig8b.eps}}}
\caption{\label{fig8} (color online) (a) to (d): Conductance as function of gate potential,
$V_g=V_{gA}=V_{gB}=V_{gC}$, obtained with FUSBMF [`+' signs (red)] and ECA (6 sites) [stars (blue)] for
several values of $t_4$. Panels (a), (b), (c) and (d) show how the peak
structure evolves from three to two peaks as the system goes from isosceles
to equilateral triangle symmetry [$t_4=t_3=0.4$, panel (d)]. For all panels,
$U=0.5$, $t_2=t_1=0.2$, and $t_3=0.4$.  
Panels (e) to (g) show ECA (6 sites) results for the partial conductances ${\rm G_i}$
(${\rm G}_1$, thick solid (red) curve; ${\rm G}_2$, solid (blue) dots curve; ${\rm G}_3$, open squares (green) curve), together
with the total conductance ${\rm G_T}$ (thin solid (black) curve). 
Panel (h) shows the evolution with $t_4$ of the phase difference $\Delta \phi_{12}$ between molecular 
orbitals $\mid\psi_1\rangle$ and $\mid\psi_2\rangle$, in units of $\pi$.}
\end{figure}

\subsection{TQD in series ($t_4=0$)}

Before presenting the results, we should point out that 
most of the embedded cluster results in this section were obtained with ECA, not LDECA.\cite{note-ldos} 
As will be seen below, in contrast to Fig.~\ref{fig2}(d), 
where the LDECA results suffer from minor finite-size effects, no 
such effects were detected after the third lead is connected. The 
higher symmetry obtained when $t_2=t_1$ leads to no discernible
finite-size effects for $t_4=t_3$. 
There are two main reasons for that. First, the third contact provides a way for the
`frozen' spin (present for $t_2=0$)\cite{finite} to delocalize from QD B. Second, when $t_4=t_3$ {\it and} $t_2=t_1$, 
any net spin would be equally distributed among the three QDs, diminishing its ability
to suppress the Kondo effect in an ECA calculation.\cite{Fabian} 

Let us start by turning on the connection of QD B to lead P, 
by varying $t_2$ from zero to $t_1=0.2$. To 
facilitate the analysis, we start with $t_4=0$ (three QDs in series, see Fig.\ \ref{fig7}). 
In this case, the molecular orbitals are: 
\begin{subequations}
\begin{eqnarray}
\mid\psi_1\rangle &=& [1,-\sqrt{2},1]/2 \\
\mid\psi_2\rangle &=& [-1,0,1]/\sqrt{2} \\
\mid\psi_3\rangle &=& [1,\sqrt{2},1]/2,
\end{eqnarray}
\end{subequations}
with $E_1=V_g-\sqrt{2}t_3$, $E_2=V_g$, and $E_3=V_g+\sqrt{2}t_3$. 
Since state $\mid\psi_2\rangle$ does not involve QD B, the 
processes of wave-splitting and loss of amplitude of the 
propagating wave through lead P will not occur when $\mid\psi_2\rangle$ 
is the state near the Fermi energy ({\it i.e.}, $V_g=-U/2$).  This results in  
the partial conductance ${\rm G}_2$ through level $\mid\psi_2\rangle$ 
being unitary, {\it i.e.}, ${\rm G}_2=G_0=2e^2/h$, 
for any value of $t_2$. This is clearly what happens to the 
central peak in Fig.~\ref{fig7}, which is associated to 
the molecular orbital $\mid\psi_2\rangle$, as previously discussed in Figs.~\ref{fig2}(a) 
and \ref{fig3}(b). 
The conductance through the other two molecular levels (rightmost and leftmost peaks in Fig.~\ref{fig7}), 
as mentioned above, will depend on the weight of QD B in $\mid\psi_1\rangle$ and 
$\mid\psi_3\rangle$. Note that, as mentioned above, because of the choice of 
parameters ($t_3/t_1=2$), the total conductance is basically the direct sum of the 
partial conductances when $t_4=0.0$ (as there is minimal overlap 
between the molecular orbitals). For $t_4=0$ (see Eqs. 25 above), 
QD B has the same coefficient in $\mid\psi_1\rangle$ and $\mid\psi_3\rangle$, 
therefore ${\rm G}_1={\rm G}_3$ for any value of $t_2$ (see the identical leftmost and rightmost peaks in Fig.~\ref{fig7}). 
The simultaneous and drastic decrease of ${\rm G}_1$ and ${\rm G}_3$ as $t_2$ increases comes 
from the increase of the coupling of QD B to lead P, which increases the 
amplitude loss through the third lead. Note the agreement between 
FUSBMF (top panel in Fig.~\ref{fig7}), and ECA (bottom panel).\cite{noteconv}

\subsection{Finite $t_4$}

An interesting picture emerges for finite $t_4$. 
In Fig.~\ref{fig8}, the top 4 panels show a comparison of conductance 
results calculated with FUSBMF (`+' (red) signs) and ECA [stars (blue)] 
for $t_4=0.1$, $0.2$, $0.3$, and $0.4$. The most salient feature in these results 
is the abrupt suppression of conductance when $t_4$ varies from 0.3 to 0.4. 
An explanation of this abrupt suppression is presented in three of the lower 
panels, which show ECA results
for the total conductance (for $t_4=0.1$, $0.3$, and $0.4$, in panels (e), (f), and (g), respectively), 
as well as partial conductances through each molecular orbital. 
In addition, panel (h) shows phase difference results [see Eq.~(24)] for paths going 
through either molecular orbital $\mid\psi_1\rangle$ or $\mid\psi_2\rangle$, for varying values of $t_4$. 

As $t_4$ increases, the coefficient of QD B in $\mid\psi_1\rangle$ 
{\it increases} monotonically (in absolute value), until it reaches $-2/\sqrt{6}$ for $t_4=t_3$, 
while it {\it decreases} monotonically for orbital $\mid\psi_3\rangle$, reaching 
$1/\sqrt{3}$ for $t_4=t_3$ [see Eqs.\ (6) and (7)]. In accordance to that, ${\rm G}_1$ {\it decreases} as 
$t_4$ increases, while ${\rm G}_3$ {\it increases}, because of the associated 
changes in the coupling of states $\mid\psi_1\rangle$ and $\mid\psi_3\rangle$ to lead P: 
more coupling ($\mid\psi_1\rangle$), more `leakage'; less coupling ($\mid\psi_3\rangle$), less `leakage'. Note that, in panels (e) 
to (g) in Fig.~\ref{fig8}, this is evident, as the solid (red) curve corresponds to 
${\rm G}_1$ and the open squares (green) curve corresponds to ${\rm G}_3$. 
In addition, since molecular orbital $\mid\psi_2\rangle$ is
independent of $t_4$ [see Eq.\ 6(b)], the maximum value of ${\rm G}_2$ (solid (blue) dots) 
is $G_0$ for all values of $t_4$ (no coupling of lead P to $\mid\psi_2\rangle$ results in no `leakage').
This can also be clearly seen in panels (e) to (g) of Fig.~\ref{fig8}, where
results for ${\rm G}_2$ are shown with solid dots (blue).
Finally, as $t_4$ approaches $t_3$, molecular orbitals $\mid\psi_1\rangle$ and $\mid\psi_2\rangle$ approach 
each other [see Fig.~\ref{fig1}(b)], allowing interference between them to 
strongly influence the transport properties of the system for $V_g$ values where 
these orbitals are close to the Fermi energy. It so happens that the phase 
difference between the paths through these two orbitals [as calculated in accordance to 
section III, Eq. (24), and shown in Fig.~\ref{fig8}(h)], for the relevant values of $V_g$, 
changes from approximately zero (for $t_4=0.1$, dashed (red) curve) to 
$\pi$ (for $t_4=t_3=0.4$, double-dot-dash (black) curve). As discussed 
in section III, the interference 
will have noticeable effects only when the molecular orbitals $\mid\psi_1\rangle$ and $\mid\psi_2\rangle$ are 
close enough in energy (for $t_4 \approx t_3$). At this point, the interference 
will be mostly destructive, as the phase difference is $\approx \pi$ 
(the resulting {\it total} conductance ${\rm G_T}$ is shown for all values of $t_4$ in panels
(e) to (g) as a thin solid (black) curve). We see the abrupt suppression of the central peak 
(associated to ${\rm G}_2$), due to its interference 
with the right-side peak (associated to ${\rm G}_1$) for $t_4=t_3$ (see panel (g) in Fig.~\ref{fig8}). 

One may ask why the suppression of the conductance in Fig.~\ref{fig8}(d) is 
less severe than the one in Fig.~\ref{fig2}(d) [note that there are no 
Fano anti-resonances in Fig.~\ref{fig8}(d)]. The reason 
is that, because of the presence of lead P, ${\rm G}_1$ is considerably less 
than the unitary conductance value $G_0$ [see thick solid (red) curve in Fig.~\ref{fig8}(g)]. 
Therefore, destructive interference cannot be total (even if the phase 
difference is $\pi$), as ${\rm G}_2$ and ${\rm G}_1$ have widely different values. 
This is not the case in Fig.~\ref{fig4}, where both partial conductances have 
similar values (being exactly the same at one $V_g$ value), as in that case lead P is not present.

The results just described for the conductance of the central peak ($\mid\psi_2\rangle$) in Fig.\ \ref{fig7}, where 
$t_2$ takes values $0$, $0.1$, and $0.2$, can also be understood in terms of the LDECA density 
of states. Figure \ref{fig9} shows the LDOS for QDs A and C [panel (a)], and QD B [panel (b)], 
for $V_g=-U/2$ (corresponding to the central peak in Fig.~\ref{fig7}). 
Notice that there is no sizable change in the value of the 
density of states at the Fermi energy ($\omega=0$), for any of the QDs, as $t_2$ varies. 
Since {\it in this case} the conductance is directly proportional to the density of states 
at the Fermi energy, this leads to a central peak in Fig.~\ref{fig7} that does 
not change with $t_2$. Obviously, this independence 
from $t_2$ comes from the fact that the density of states of QD B is very small in a broad interval 
around the Fermi energy when the charge transport occurs through orbital $\mid\psi_2\rangle$, 
resulting in lead P being effectively disconnected from the TQD for this value of $V_g$. 
Notice that as lead P couples to the other molecular orbitals ($\mid\psi_1\rangle$ and $\mid\psi_3\rangle$), there is no longer a simple proportionality relation 
between the LDOS and the conductance,\cite{leadP} so that no simple direct connection 
can be made between the LDOS at the Fermi energy and the 
conductance (as was done in Fig.~\ref{fig3}). 

\begin{figure}
\centerline{\resizebox{3.5in}{!}{\includegraphics{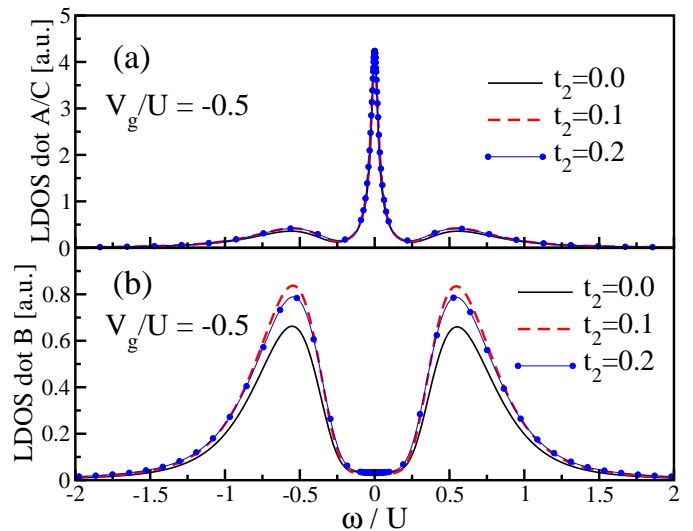}}}
\caption{ (color online) LDOS calculated using LDECA (6 sites) for $U=0.5$, $t_1=0.2$, $t_3=0.4$, $t_4=0$, and varying values of $t_2$,
at the particle-hole symmetric point ($V_g/U=-0.5$). Panel (a) contains results for 
QDs A and C and panel (b) for QD B. 
In panel (a), it is evident that coupling QD B to lead P ($t_2=0.1$, dashed (red) curve, 
and $t_2=0.2$, solid (blue) dots curve) has no effect over the density of states 
of QDs A and C, as curves are indistinguishable from the $t_2=0$ case
(solid (black) curve). Thus, the conductance does 
not change with $t_2$ (see Fig.~\ref{fig7}). Note in panel (b) that QD B 
has no spectral weight at $\omega=0$ for any value of $t_2$, implying that 
when charge transport occurs through orbital $\mid\psi_2\rangle$, the conductance of 
the TQD is not affected by its coupling to lead P.
}
\label{fig9}
\end{figure}

\section{Conclusions}

We have studied the transport properties of a TQD in the molecular regime coupled to
leads. By applying the FUSBMF, and the ECA and LDECA approaches, we have calculated the conductance of the TQD 
system for different symmetries and different configurations of the leads. For the two-leads case, we have
calculated the conductance for both series and triangle
configurations. In the series configuration, LDECA and FUSBMF results agree with 
each other and with the results for the molecular regime 
obtained in Ref.~\onlinecite{zitko1}, where the Kondo effect has been studied in
detail. In the triangular symmetry, 
the quantitative results obtained by FUSBMF and LDECA differ slightly as the 
equilateral symmetry is approached, due to a finite-size effect 
in the relatively small clusters accessible to LDECA, although 
agreement is still very good [see Fig.~\ref{fig2}(d)]. The suppression of conductance in the regime 
where approximately two electrons occupy the triangle was explained by LDECA as an 
interference effect between two degenerate molecular orbitals, utilizing the concept 
of partial conductance. In addition, our results for triangular 
symmetry differ from those presented recently in Ref.~\onlinecite{lobos}. 
We believe that the approach pursued here, where details of the 
internal structure of the interacting region of the system are taken fully into
account, are very important to explain the conductance of the TQD system. In fact, our results show that changes in the internal couplings of the TQD dramatically change the features of
the conductance. We also found that the degeneracy of the molecular orbitals 
at equilateral symmetry, when two electrons occupy the TQD, induces an effective 
ferromagnetic interaction between the spins localized in the interacting region,\cite{ferro} 
leading to an $S=1$ Kondo effect.

In the TQD series configuration, our results show that the third lead produces a
strong suppression in the bonding and anti-bonding orbital conductance peaks 
(Fig.~\ref{fig7}). The non-bonding peak, however, remains 
unchanged, since this orbital does not have the appropriate symmetry to couple to
lead P.\@ This suppression of conductance can be seen as a `loss of 
amplitude' through lead P, similar to the effect occurring with beam splitters in optics. 
If one thinks of the conductance in terms of transmission of waves through the 
interacting region, the introduction of lead P provides an additional 
transmission channel, which clearly affects the conductance 
between leads L and R. This `loss of amplitude' idea is then used to understand the 
conductance results in the triangular symmetry. In particular, 
it explains why the interference effects seem less effective 
in suppressing the conductance in the equilateral symmetry ({\it i.e.}, 
why no Fano anti-resonance occurs): the `loss of amplitude' prevents the conductance 
through molecular orbital $\mid\psi_1\rangle$ 
from reaching the unitary limit, leading to a decrease in the destructive interference, 
as discussed in detail in Fig.~\ref{fig8}. 
We should remark that the excellent overall quantitative agreement of results 
obtained with FUSBMF, ECA, and LDECA\cite{finite,noteconv} (which
rely on totally different approximations) makes our conclusions much more reliable and robust.  
Moreover, the combination of techniques allows a better insight into the physics of the different geometries.  

\begin{acknowledgements}
The authors wish to acknowledge fruitful discussions with E. H. Kim and K. Ingersent.
E.V.A. thanks the Brazilian agencies FAPERJ, CNPq (CIAM project), and CAPES for financial support.  Work at Ohio was partially supported by NSF grant DMR-0710581; at Oakland it was supported by NSF grant DMR-0710529. 
\end{acknowledgements}


\begin{thebibliography}{apssamp}
\bibitem{note-molecular} Although some of the results presented in section III 
are for parameter values such that $t_3 \approx t_1$, we checked that 
the results are qualitatively similar to the ones obtained for $t_3 \gg t_1$. 
Therefore, one can be confident that the parameters used in section III 
reflect the molecular regime situation (see also section III-E).
\bibitem{g0} Note that the partial conductances ${\rm G}_1$ and ${\rm G}_2$ in 
Fig.~\ref{fig4} have maximum values larger than the conductance quantum ($G_0=2e^2/h$). 
As alluded above, the partial conductances, when their associated molecular orbitals 
overlap with each other, cannot be independently measured, having therefore no 
physical significance by themselves, and therefore do not have to be $\leq G_0$. 
When the molecular orbital levels are well separated in energy 
(and therefore the partial conductances match the total (measured) conductance), then 
${\rm G_i} \leq G_0~\forall~{\rm i}$. For all cases, moreover, the total conductance 
${\rm G_T} \leq G_0$, as one expects on physical grounds.
\bibitem{finite} The origin of the finite size effect is an unfavorable
geometry for small clusters for $t_3 \approx t_4$, which tends to `freeze' a spin in QD B.
This suppresses spin fluctuations, with a consequent suppression of the Kondo effect.
For a comprehensive discussion of similar effects in different systems (using
ECA and DMRG), see Ref.~\onlinecite{Fabian}.
\bibitem{note-double} Note that results for $\langle n_i \rangle$ in Fig.~\ref{fig5} 
have to be multiplied by 2 to account for different spin orientations.
\bibitem{note-ldos} Except, obviously, for the LDOS results, where LDECA was used with $\lambda \gtrsim 1.0$.
\bibitem{noteconv} Although there is good qualitative agreement between FUSBMF and ECA in Fig.\ \ref{fig7}, the ECA conductance peaks are narrower. 
This arises from the fact that the clusters used for these calculations had just 
6 sites, and therefore the results are not fully size-converged. The next cluster size 
available for this geometry contains 12 sites. Although feasible, ECA calculations for this 
cluster size consume considerably more CPU time.
\bibitem{leadP} To confirm this, a test for a simplified system containing a single-level 
QD with three leads shows that the conductance through two of the leads is not directly 
proportional to the LDOS at the QD.
\end{thebibliography}
\end{document}